\documentclass{article}

\usepackage{PRIMEarxiv}

\usepackage[utf8]{inputenc} 
\usepackage[T1]{fontenc}    
\usepackage{hyperref}       
\usepackage{url}            
\usepackage{booktabs}       
\usepackage{amsfonts}       
\usepackage{nicefrac}       
\usepackage{microtype}      
\usepackage{lipsum}
\usepackage{fancyhdr}       
\usepackage{graphicx}       
\graphicspath{{media/}}     
\usepackage{tabularx}
\usepackage{multirow}
\usepackage{makecell}
\usepackage{caption}
\usepackage{subcaption}
\usepackage{multirow}
\usepackage{colortbl}
\usepackage{float}
\usepackage{amsmath}
\title{Effects of Using Synthetic Data on Deep Recommender Models' Performance
}

\author{
 Fatih Cihan Taskin \\
  AI Enablement \\
  Huawei Türkiye R\&D Center \\
  Istanbul, Turkey \\
  \texttt{fatih.cihan.taskin@huawei.com} \\
   \And
 Ilknur Akcay \\
  AI Enablement \\
  Huawei Türkiye R\&D Center \\
  Istanbul, Turkey\\
  \texttt{ilknur.akcay@huawei.com} \\
  \And
 Muhammed Pesen \\
  AI Enablement \\
  Huawei Türkiye R\&D Center \\
  Istanbul, Turkey \\
  \texttt{muhammedpesenn@gmail.com} \\
  \And
\hspace{6mm}Said Aldemir \\
\hspace{6mm} AI Enablement \\
\hspace{9mm} Huawei Türkiye R\&D Center \\
\hspace{6mm} Istanbul, Turkey \\
\hspace{6mm}  \texttt{said.aldemir1@huawei.com} \\
  \And
\hspace{5mm}  Ipek Iraz Esin \\
\hspace{5mm}   AI Enablement \\
\hspace{5mm}   Huawei Türkiye R\&D Center \\
\hspace{5mm}   Istanbul, Turkey \\
\hspace{5mm}   \texttt{ipek.iraz.esin@huawei.com} \\
  \And
Furkan Durmus \\
AI Enablement \\
Huawei Türkiye R\&D Center \\
Istanbul, Turkey \\
\texttt{furkan.durmus2@huawei.com}
  \And
}

\begin{document}

\maketitle

\begin{abstract}
Recommender systems are essential for enhancing user experiences by suggesting items based on individual preferences. However, these systems frequently face the challenge of data imbalance, characterized by a predominance of negative interactions over positive ones. This imbalance can result in biased recommendations favoring popular items. This study investigates the effectiveness of synthetic data generation in addressing data imbalances within recommender systems. Six different methods were used to generate synthetic data. Our experimental approach involved generating synthetic data using these methods and integrating the generated samples into the original dataset. Our results show that the inclusion of generated negative samples consistently improves the Area Under the Curve (AUC) scores. The significant impact of synthetic negative samples highlights the potential of data augmentation strategies to address issues of data sparsity and imbalance, ultimately leading to improved performance of recommender systems.
\end{abstract}


\section{Introduction}

Recommender systems have become significant components of digital advertising platforms that recommend the most suitable ads to users. Especially with the development of deep learning algorithms, modeling complex user behaviors has been attempted to increase both the revenue of companies and improve user satisfaction by enhancing the success of recommender models \cite{Zhang_2019}. The main goal of a recommender model is the accurate prediction of Click-Through Rates (CTR), which measure the likelihood of a user interacting with a given item. Since the click rate of users on the displayed advertisements is generally low \cite{richardson_2007}, the dataset used in training the models will be quite imbalanced. This imbalance poses substantial challenges like popularity bias, user activity bias, and poor generalization, hence lower recommendation quality \cite{wei_2021, chen_2023}. Disparities in user activity and item popularity can negatively affect a model's learning process, creating a self-reinforcing cycle where already popular items gain even more prominence. This phenomenon, known as a feedback loop, prioritizes the preferences of highly active users while overlooking those of less engaged individuals  \cite{mansoury2020feedback}. As a result, the user experience suffers due to increased homogeneity in recommendations, which limits exposure to diverse content. Moreover, this loop hinders the visibility of niche items, ultimately undermining the long-term objectives of promoting diversity and ensuring fairness within the recommendation system.

To address the data imbalance problem,  two main approaches are typically employed: resampling strategies and weighting mechanisms \cite{Imbalancedlearning}. Resampling techniques, such as oversampling the minority class or undersampling the majority class, aim to equalize the class distribution by manipulating the training data. On the other hand, weighting mechanisms assign different weights to the samples depending on the class so that the minority class becomes more important during the training process. Although traditional methods have been effective in mitigating the impact of unbalanced datasets, the use of synthetic data generation has gained popularity in recent years \cite{fonseca2023tabular} due to its ability to create large, diverse, and balanced datasets. This approach not only helps improve the performance of machine learning models but also addresses limitations such as data scarcity and privacy concerns. Simple oversampling methods and generative models are mostly used to create artificial samples that resemble the characteristics of the minority class. The imbalance between the classes can be mitigated by augmenting the training data with these synthetically generated samples, leading to improved model performance and generalization.

The use of synthetic data generation techniques is primarily associated with data privacy concerns in deep learning context \cite{lu2024machine}. By creating artificial data that mimics the characteristics of the original dataset, companies can ensure the anonymity of sensitive information while still conducting meaningful analysis. This approach has proven useful in maintaining confidentiality and mitigating potential risks associated with data privacy concerns. However, in this study, we pursue the research question of how the success of CTR prediction tasks is affected by synthetic data generation. 

In our study, we employed a variety of techniques to generate synthetic datasets to augment our original data. The methods used include random oversampling, several variants of the Synthetic Minority Over-sampling Technique (SMOTE) \cite{smote_2002}, Conditional Tabular Generative Adversarial Network (CTGAN) \cite{CTGan_2019}, Gaussian Copula \cite{SDV_2016}, Copula Generative Adversarial Network (Copula GAN) \cite{SDV_2016}, Tabular Variational AutoEncoder (TVAE) \cite{CTGan_2019}, and the Tabular Diffusion Probabilistic Model (TabDDPM) \cite{TabDDPM_2022}. It aims to observe the impact of synthetic data by exploring several scenarios and a variety of techniques. In different synthetic data generation scenarios, only positive samples were produced and added to the original data at the rate of 25\% and 50\%, only negative samples were produced and added to the original data at the rate of 25\% and 50\%, and also hybrid only datasets with the same amount of original data belonging to both labels were produced and used instead of the original dataset. As a result, 5 different datasets were created and evaluated by training CTR prediction models.

Subsequently, we conducted extensive experiments to assess how the inclusion of these synthetic datasets influenced the offline performance of our deep recommender models. The aim was to determine which synthetic data generation technique and scenario enhanced the model's performance most effectively, providing insights into the optimal strategies for data augmentation in recommender systems.

\begin{figure}[h]
\centering
\label{main_figure}
\includegraphics[width=0.9\linewidth]{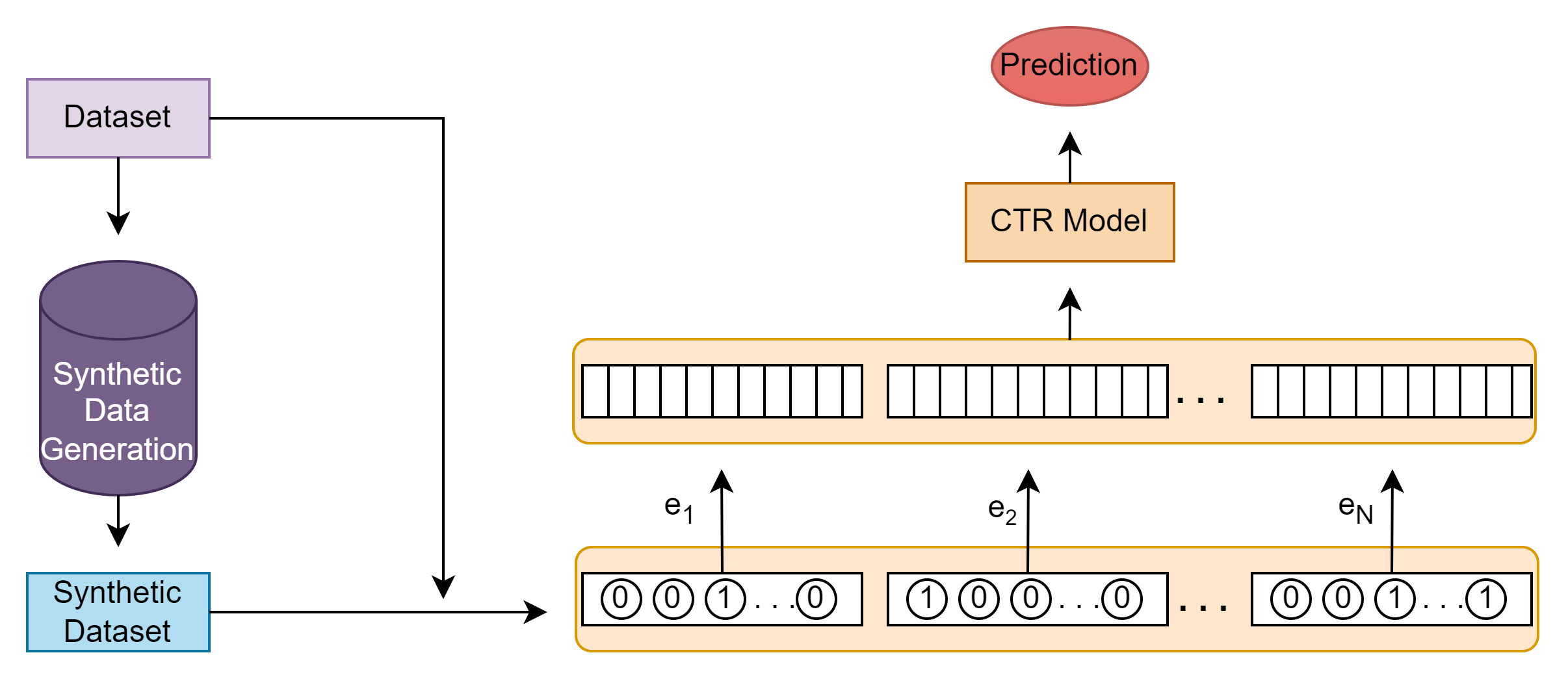}
\caption{A real dataset is augmented with synthetic data generated from the original dataset to enhance the training process. This combined dataset undergoes an embedding operation to transform sparse features to embedding vectors. Following this, the input embedding matrix is fed into the CTR prediction model, which uses the embedded information to predict the likelihood of a user clicking on an advertisement by capturing complex relationships among the features.}
\end{figure}

\section{Related Works}

In deep learning, the quality and quantity of data have a direct impact on the accuracy and generalization capability of the model. Generating synthetic data is a solution to critical problems such as data scarcity, high collection costs, and privacy concerns. By creating realistic and diverse datasets, synthetic data facilitates the development of deep learning models in various domains like computer vision and natural language processing \cite{Khader2023, medical_fundus, medical_tumor_seg, Loecher2021, Skandarani2023}. This approach not only improves the performance of the models but also enables the inclusion of rare and unusual scenarios, ensuring better adaptability and reliability in real-world applications. While GANs and VAEs are generally used previously, diffusion models have become widespread as an alternative to GANs in recent years and have shown success in many different fields \cite{Ho_2020}. It is claimed that diffusion models are more successful compared to GANs, but problems such as diffusion models being more data greedy and more costly are also mentioned \cite{diffusion_vs_gans, UsmanAkbar2024}.

Vallelado et al. proposed the use of diffusion models with transformer conditioning for data imputation and generation \cite{villaiz2024diffusion}. Diffusion models, known for capturing complex data distributions, are enhanced with transformers to model dependencies and long-range interactions within tabular data. Their approach was evaluated against state-of-the-art techniques such as Variational Autoencoders (VAEs) and Generative Adversarial Networks (GANs) on benchmark datasets. For data imputation, the models' accuracy in estimating missing values while preserving data distribution was assessed. For data generation, the quality and diversity of synthetic data samples were evaluated.

Endres et al. conducted a comparative study on 3 tabular datasets by using GAN-based, SMOTE, and VAE-based models \cite{Endres2022}. The produced datasets were evaluated by statistical metrics and processing time. However, the generated datasets were not used in deep learning model training. Additionally, they did not consider the numeric and categorical features, especially for the SMOTE method.

Slokom et al. proposed SynRec, which is a data protection framework that uses data synthesis to protect sensitive information in recommender systems \cite{synrec_2018}. It replaces original values with synthetic ones or generates new users, ensuring that sensitive information is concealed while maintaining data usability for comparing recommender systems. This enables companies to safely share data with researchers for algorithm development and collaborative research. The paper demonstrates feasibility of the concept through preliminary experiments and outlines future challenges for practical implementation.

Noble et al. proposed two metrics, Identifiability (measuring privacy risk) and Realism (comparing recommendation performance between real and synthetic data), because there's a trade-off between the detailed user data needed for high-quality recommendations and privacy concerns \cite{Noble2023RealisticBN}. Analyzing seven data generation algorithms for movies and songs across 28 settings, they constructed Pareto frontiers of Realism vs. Identifiability, offering insights to guide future synthetic data research.

\section{Synthetic Data Generation Methods}

\textbf{SMOTEN}:
SMOTE (Synthetic Minority Over-sampling Technique) is a synthetic data generation method that aims to balance of imbalanced datasets. To generate data, k-nearest neighbors are found for each sample in the minority class. While these neighbors can typically be found using the Euclidean distance, they can also be found using different distance metrics. The action is taken according to the number of neighbors specified in the parameter. Synthetic data is produced according to interpolation between the sample data and the selected neighbor. Different SMOTE algorithms can be used depending on the type of dataset used and the selection of neighborhoods. SMOTE algorithm works with numeric datasets. There are also SMOTE variants that work with different data types. In this study, the SMOTEN method (Synthetic Minority Over-sampling Technique for Nominal Features) is specifically tailored to handle categorical data by creating synthetic samples through a process that respects the nominal nature of the features. To address memory constraints, we selected a small value for the k-neighbors parameter, setting it to 2. This choice helped to manage computational resources effectively. In addition, the dataset was divided into 4 chunks containing around 50000 samples each due to memory constraints. The SMOTEN algorithm is run for each chunk, and finally, all chunks are merged.

\textbf{CT-GAN}:
CTGAN is a generative adversarial network (GAN) architecture designed to generate synthetic tabular data. CTGAN focuses on learning the joint distribution of the data through adversarial training, aiming to generate structured data that closely resembles the statistical properties of the original dataset in terms of feature correlations and distributions. It consists of two main components: a generator and a discriminator. The generator takes random noise as input and transforms it into synthetic data points. It generates synthetic data that resembles the statistical properties in the original dataset. The discriminator aims to distinguish between samples from the original dataset and synthetic samples produced by the generator and provides feedback to the generator on how realistic its generated samples are. Since these two components are engaged in an adversarial training process, while the discriminator improves at identifying synthetic data, the generator generates gradually higher-quality synthetic data. CTGAN is beneficial for applications that need synthetic data generation, such as data augmentation, privacy-preserving data sharing, and machine learning model testing. For the model structure, the embedding dimension was 128, with both the generator and discriminator dimensions set to (256, 256). The learning rates for both the generator and discriminator were 0.0002, with a decay of 1e-06 for each. The batch size was 500, and the discriminator steps were set to 1. Logging frequency was enabled, while verbose mode was disabled. The number of epochs was set to 300, with a pack size (PAC) of 10. CUDA was enabled for computation.

\textbf{CopulaGAN}:
CopulaGAN is a generative adversarial network (GAN) architecture designed to generate synthetic tabular data that preserves complex relationships between variables. CopulaGAN uses copulas to describe the dependence structure between random variables while separately modeling their marginal distributions. With this concept, synthetic data that preserves the marginal distributions and dependencies observed in the original dataset is generated. Unlike CTGAN, CopulaGAN focuses on capturing the joint distribution of variables more explicitly through the use of copulas. Since CopulaGAN separates the modeling of marginal distributions and dependencies, it effectively captures complex relationships in tabular data, leading to more realistic and convenient synthetic datasets compared to traditional GAN approaches. CopulaGAN, like CTGAN, is beneficial for applications that need synthetic data generation, such as data augmentation, privacy-preserving data sharing, and machine learning model testing. For the model structure, the embedding dimension was 128, with both the generator and discriminator dimensions set to (256, 256). The learning rates for both the generator and discriminator were 0.0002, with a decay of 1e-06 for each. The batch size was 500, and the discriminator steps were set to 1. Logging frequency was enabled, while verbose mode was disabled. The number of epochs was set to 300, with a pack size (PAC) of 10. CUDA was enabled for computation.

\textbf{TVAE}:
The Tabular Variational Autoencoder (TVAE) model was developed to produce artificial tabular data. By utilizing the concepts of variational autoencoders (VAEs), TVAE is able to provide synthetic data that closely resembles the statistical characteristics and patterns found in the original dataset by capturing both the joint and conditional distributions of data in a tabular format. The architecture of the model is comprised of two parts: an encoder that converts actual tabular data into a latent space representation and a decoder that uses this latent representation to reconstruct the tabular data. TVAE generates accurate and diverse synthetic data by using a specialized loss function that combines Kullback-Leibler (KL) divergence and reconstruction loss. For the model structure, the embedding dimension was 128, with both the compress and decompress dimensions set to (128, 128). The L2 regularization parameters were 1e-06. The batch size was 500. The number of epochs was set to 300. CUDA was enabled for computation.

\textbf{TabDDPM}:
Diffusion models are one of the relatively new classes of generative models that have shown promise in addressing class imbalances in datasets. By employing a technique similar to reverse Markov chains, they gradually convert random noise into samples that resemble the targeted collection of data. These models produce complex and varied synthetic samples that are similar to real data through iterative adjustments.
Diffusion models are particularly useful because they can generate high-quality synthetic examples. They can significantly enhance the representation of insufficiently represented classes in the training data through the generation of synthetic samples. These models can improve the performance of machine learning tasks and effectively address class imbalances when included in the data preprocessing pipeline. The model structure used in this study includes a multi-layer perceptron (MLP) with layers [512, 1024, 1024, 1024, 1024, 1024, 512] and no dropout. The diffusion process is conducted over 100 timesteps with a Gaussian loss type set to "mse" and a scheduler set to "cosine." For training, normalization is performed using the quantile method.

\textbf{Gaussian Copula}:
Gaussian Copula treats each column in data as a variable and uses multivariate probability distribution models to determine the relationship between variables. Firstly, the marginal distribution of each variable is determined. Then, multivariate normal distribution is used to see the relationship of the variables with each other. Gaussian Copula combines these two distributions to establish a relationship between the data. This allows us to capture linear dependencies between variables. For the model structure, numerical distributions was None and default distributions was 'beta'.

\section{Experimental Setup}

To evaluate the impact of synthetic data on recommendation models, we utilized the Frappe\footnote{https://www.baltrunas.info/context-aware} dataset. Additionally, we used FuxiCTR\footnote{https://fuxictr.github.io/tutorials/v2.0/} repository \cite{fuxi_2021, fuxi_2022} due to its ability to ensure the repeatability and reliability of our experiments, which is crucial for validating our findings. The experimental environment was configured with the following specifications: Operating System: Ubuntu 22.04.3 LTS, Python Version: 3.9.7, Python Distribution: Anaconda3, RAM: 252 Gb, CPU: Intel(R) Xeon(R) CPU E5-2650 v4 @ 2.20GHz, and GPU: NVIDIA TITAN RTX. 

Five distinct scenarios were designed to systematically observe the effects of synthetic data on the recommendation models. For each scenario, it is ensured that only the training data is changed according to the synthetic data generation scenario, and the validation and test sets are kept unchanged to maintain unbiased results. By isolating the changes to just the training data, we can accurately assess how those specific modifications impact the model's performance on unseen data. Keeping the validation and test sets consistent across all scenarios allows for a fair comparison of the different synthetic data generation techniques. These scenarios are described as follows:

\textbf{Scenario 1 (S1)}: In this scenario, synthetic positive samples amounting to 25\% of the original number of positive samples in the dataset were generated. These synthetic samples were then added to the original dataset. The purpose was to observe the effect of a modest increase in positive interactions on the model's performance. 

\textbf{Scenario 2 (S2)}: Here, synthetic positive samples equal to 50\% of the original positive samples were created and incorporated into the original dataset. This scenario aimed to investigate the impact of a more substantial addition of positive data on the recommendation model. 

\textbf{Scenario 3 (S3)}: For this scenario, we introduced synthetic negative samples equivalent to 25\% of the original number of negative samples. These were added to the dataset to understand how a slight increase in negative interactions influences the model's recommendations. 

\textbf{Scenario 4 (S4)}: In this case, synthetic negative samples amounting to 50\% of the original negative samples were generated and added to the dataset. This scenario was designed to evaluate the effects of a significant increase in negative data on the model's accuracy and reliability. 

\textbf{Scenario 5 (S5)}: This scenario involved generating a completely synthetic dataset that matched the original dataset in size. The aim was to assess the performance of the recommendation models when trained on entirely synthetic data, comparing it directly with models trained on the original dataset in terms of the CTR prediction performance. When the SMOTE algorithm generates data, it creates a new dataset by adding synthetic data over the original data. Therefore, when running the S5 scenario with the SMOTE algorithm, firstly generated a dataset that is approximately 2 times the size of the existing dataset. The generated dataset contained both the original data and the synthetic data. The original data was removed from this dataset, and only the synthetic data was used.

In scenarios S1, S2, S3, and S4, we could not create synthetic data that had only positive or negative samples. Instead, we generated a hybrid dataset in S5 that consisted of only synthetic data and then extracted the positive and negative samples from this hybrid dataset to incorporate the desired proportions into the original dataset. By comparing these scenarios, we aimed to thoroughly understand how different proportions and types of synthetic data affect the performance of recommendation models.

Since the presence of some feature combinations during synthetic data production disrupts the data distribution, a constraint has been introduced for these combinations. For example, in the country and city features in one sample, the country is Brazil, but the city is Istanbul, which disrupts the distribution of the data and increases noise. To prevent this, a feature-by-feature analysis was performed to analyze which features had unique values and which did not, and the constraint was used in CTGAN, TVAE, CopulaGAN, and GaussianCopula methods to prevent combinations that were not in the original data from occurring in synthetic data.

\begin{table}[h]
\caption{This table shows the number of samples that identically match between the synthetic dataset and original dataset for each method.}
\label{identical_samples_table}
\begin{tabular}{ccccccc}
                  & Gaussian Copula & TVAE & CTGan & CopulaGan & Diffusion & Smote \\ \hline
Identical Samples & 30       & 1226 & 489   & 292       & 755       & 41272 \\ \hline
\end{tabular}
\end{table}

Figure 1 shows the main pipeline of synthetic data generation and training of CTR prediction models. A new dataset is created by combining synthetic data derived from the original dataset to improve the model training process. This combined dataset is then subjected to an embedding operation, transforming sparse features into embedding vectors. Subsequently, the resultant input embedding matrix is fed into the CTR prediction model, which leverages the embedded information to predict the likelihood of a user clicking on an advertisement by capturing complex interactions among the features. 3 different recommender models are chosen: DNN, DeepFM, and Masknet which are described briefly below.

\textbf{DNN}: Deep Neural Network (DNN) is the baseline CTR prediction model that consists of multiple hidden layers between the input and output layers. It is used to learn complex patterns and interactions from sparse feature inputs in CTR prediction problems.

\textbf{DeepFM} \cite{deepfm}: DeepFM is a hybrid model combining the strengths of Factorization Machines (FM) and DNNs in the CTR prediction task. The FM component is responsible for modeling low-order feature interactions, whereas the DNN component learns high-order feature interactions through multiple hidden layers. The outputs of the FM component and DNN component are combined through concatenation.

\textbf{MaskNet} \cite{masknet}: MaskNet model proposed a novel feature masking mechanism to perform an adaptive selection of the most relevant features for each input.

\begin{table}[h!]
    
\caption{Performance comparison of CTR prediction models trained on datasets from 5 different scenarios, produced using 6 different synthetic methods in terms of AUC score. For each synthetic data generation scenario, 3 different CTR prediction models were trained. Bold values indicate the highest results among different scenarios for each model.}
\label{results_table}
\centering
\scalebox{0.965}{%
\begin{tabular}{lccccccccc}
\hline
\multirow{2}{*}{\textbf{Scenario}} & \multicolumn{3}{c}{\textbf{GaussianCopula}}                        & \multicolumn{3}{c}{\textbf{TVAE}}                                  & \multicolumn{3}{c}{\textbf{CTGAN}}                                 \\ \cline{2-10} 
                                   & \textbf{DeepFM}      & \textbf{DNN}         & \textbf{MaskNet}     & \textbf{DeepFM}      & \textbf{DNN}         & \textbf{MaskNet}     & \textbf{DeepFM}      & \textbf{DNN}         & \textbf{MaskNet}     \\ \hline
Original                               & 0.98409              & 0.98396              & 0.98329              & 0.98409              & 0.98396              & 0.98329              & 0.98409              & 0.98396              & 0.98329              \\ \hline
S1                                 & 0.98233              & 0.98245              & 0.98131              & 0.98307              & 0.98216              & 0.98144              & 0.98244              & 0.98163              & 0.98098              \\ \hline
S2                                 & 0.98231              & 0.98192              & 0.98143              & 0.98191              & 0.98247              & 0.98061              & 0.98213              & 0.98172              & 0.98157              \\ \hline
S3                                 & \textbf{0.98480}     & \textbf{0.98510}     & \textbf{0.98447}     & \textbf{0.98417}     & \textbf{0.98456}     & \textbf{0.98403}     & 0.98454              & 0.98420              & \textbf{0.98395}     \\ \hline
S4                                 & 0.98461              & 0.98419              & 0.98438              & 0.98389              & 0.98332              & 0.98284              & \textbf{0.98461}     & \textbf{0.98485}     & 0.98384              \\ \hline
S5                                 & 0.78847              & 0.78935              & 0.79928              & 0.73345              & 0.73176              & 0.73352              & 0.78643              & 0.78876              & 0.79026              \\ \hline
                                   & \multicolumn{1}{l}{} & \multicolumn{1}{l}{} & \multicolumn{1}{l}{} & \multicolumn{1}{l}{} & \multicolumn{1}{l}{} & \multicolumn{1}{l}{} & \multicolumn{1}{l}{} & \multicolumn{1}{l}{} & \multicolumn{1}{l}{} \\ \hline
\multirow{2}{*}{\textbf{Scenario}} & \multicolumn{3}{c}{\textbf{CopulaGAN}}                             & \multicolumn{3}{c}{\textbf{Diffusion}}                             & \multicolumn{3}{c}{\textbf{SMOTEN}}                                 \\ \cline{2-10} 
                                   & \textbf{DeepFM}      & \textbf{DNN}         & \textbf{MaskNet}     & \textbf{DeepFM}      & \textbf{DNN}         & \textbf{MaskNet}     & \textbf{DeepFM}      & \textbf{DNN}         & \textbf{MaskNet}     \\ \hline
Original                               & 0.98409              & 0.98396              & 0.98329              & 0.98409              & \textbf{0.98396}     & 0.98329              & 0.98430              & 0.98396              & 0.98329              \\ \hline
S1                                 & 0.98118              & 0.98167              & 0.98069              & 0.97920              & 0.98178              & 0.97759              & 0.98386              & 0.98398              & 0.98316              \\ \hline
S2                                 & 0.98218              & 0.98207              & 0.98078              & 0.98061              & 0.98096              & 0.97501              & 0.98381              & 0.98357              & 0.98348              \\ \hline
S3                                 & 0.98390              & 0.98445              & 0.98320              & 0.98438              & 0.98199              & 0.98317              & 0.98387              & \textbf{0.98433}     & 0.98333              \\ \hline
S4                                 & \textbf{0.98500}     & \textbf{0.98498}     & \textbf{0.98369}     & \textbf{0.98455}     & 0.98450              & \textbf{0.98402}     & \textbf{0.98476}     & 0.98413              & \textbf{0.98399}     \\ \hline
S5                                 & 0.74251              & 0.73928              & 0.74523              & 0.74668              & 0.74287              & 0.75931              & 0.96481              & 0.96399              & 0.96287              \\ \hline
\end{tabular}}
\end{table}

\section{Discussion}

In this study, we explored the impact of synthetic data generation techniques on the performance of deep recommender systems. For this purpose, we employed 6 methods: Synthetic Minority Over-sampling Technique for Nominal features (SMOTEN), Conditional Generative Adversarial Networks (CTGAN), Copula Generative Adversarial Network (CopulaGAN), Tabular Variational Autoencoders (TVAE), Gaussian Copula and Tabular Denoising Diffusion Probabilistic Model (TabDDPM) to generate synthetic data and evaluated their influence on CTR prediction performance by comparing the Area Under Curve (AUC) metric by using synthetic and original datasets in 5 different scenarios with 3 different deep recommender models.

According to the experimental results, the AUC increased in scenarios S3 and S4, except for the DNN model when using the TabDDPM method. The dataset has a CTR rate of 0.33, meaning it is highly imbalanced with few positive samples. Despite this, generating more positive samples did not improve the AUC for any method and model. This shows that trying to balance the dataset by adding positive samples is ineffective. Interestingly, even though there were more negative samples, adding synthetic negative samples did increase the AUC score.

Scenario S5 shows the results of models trained on only synthetic data without using any original data. This scenario aims to assess how well the generated datasets match the real dataset. The key point here is that SMOTEN produced the most similar results to the original data and the highest AUC scores in S5. However, it is important to note that SMOTEN essentially copies the original dataset and produces nearly identical samples, which is not desired. If the goal of using synthetic data were to address privacy concerns, SMOTEN would fail. Higher AUC scores do not mean that SMOTEN is successful in this context. Table \ref{identical_samples_table} shows the number of identical samples within the real dataset. Clearly, it can be realized that SMOTEN copies most of the samples identically by comparing the other methods. This can be the reason why it gets higher AUC results for S5.

Overall, the Gaussian Copula method works better in generating synthetic datasets. In the S5 scenario, AUC values get higher among all methods if we exclude SMOTE due to the given reason above. Regarding the GaussianCopula model, which is the best-performing one, the original AUC values for DeepFM, DNN, and MaskNet are 98.409\%, 98.396\%, and 98.329\%, respectively. In the best-performing scenario (S3), the AUC values increase to 98.480\% for DeepFM, 98.510\% for DNN, and 98.447\% for MaskNet. This represents an AUC increase of 0.071\% for DeepFM, 0.114\% for DNN, and 0.118\% for MaskNet.

This study has limitations that open doors for further research. The experiments were likely conducted on a specific dataset and recommender system architecture. Exploring the applicability and generalizability of these findings across different datasets and model architectures would be valuable. Additionally, investigating the impact of synthetic data generation on other recommendation metrics beyond AUC, such as novelty or diversity, would provide a more comprehensive understanding of its effects. 

While our study demonstrates the potential benefits of synthetic data generation in improving deep recommender system performance, several limitations need to be addressed in future research. First, the quality of the synthetic data heavily depends on the generation method and the complexity of the original dataset. More research is needed to determine the optimal generation method for different types of datasets and recommendation tasks.

Lastly, the generalizability of our findings across different domains and datasets should be explored. Further experimentation on a broader range of datasets and recommendation scenarios will provide a more comprehensive understanding of the effectiveness and limitations of synthetic data generation in deep recommender systems.

\section{Conclusion}
 
Our findings indicate that the most significant performance improvement in terms of the AUC scores was observed when only generated negative samples were added to the original dataset. Adding only positive samples or both types of samples, whether combined with the original data or used separately, did not yield the same level of improvement.

However, it is important to note that generating synthetic datasets is a high-cost process that requires substantial computational resources and careful design. Therefore, it is crucial to consider the potential benefits and costs very carefully before implementing synthetic data generation in practical applications. Furthermore, our experiments were conducted in an offline setting, which may not fully capture the complexities and dynamics of real-world usage. To validate these findings, it is essential to conduct online experiments and A/B testing to observe the effects in a live environment. This will help to ensure that the improvements seen offline translate to actual user interactions and enhance the overall effectiveness of recommender systems.

Future work should investigate the underlying reasons why negative samples have a more substantial impact on CTR model performance and should explore different data generatio methods to further optimize synthetic data generation for recommender systems. Additionally, it would be beneficial to test these findings across different domains and datasets to ensure their generalizability.

\newpage
\bibliographystyle{unsrt}  
\bibliography{sample-base}

\begin{thebibliography}{10}

\bibitem{Zhang_2019}
Shuai Zhang, Lina Yao, Aixin Sun, and Yi~Tay.
\newblock Deep learning based recommender system: A survey and new perspectives.
\newblock {\em ACM Computing Surveys}, 52(1):1–38, February 2019.

\bibitem{richardson_2007}
Matthew Richardson, Ewa Dominowska, and Robert Ragno.
\newblock Predicting clicks: estimating the click-through rate for new ads.
\newblock In {\em Proceedings of the 16th international conference on World Wide Web}, pages 521--530, 2007.

\bibitem{wei_2021}
Tianxin Wei, Fuli Feng, Jiawei Chen, Ziwei Wu, Jinfeng Yi, and Xiangnan He.
\newblock Model-agnostic counterfactual reasoning for eliminating popularity bias in recommender system.
\newblock In {\em Proceedings of the 27th ACM SIGKDD Conference on Knowledge Discovery \& Data Mining}, KDD '21, page 1791–1800, New York, NY, USA, 2021. Association for Computing Machinery.

\bibitem{chen_2023}
Jiawei Chen, Hande Dong, Xiang Wang, Fuli Feng, Meng Wang, and Xiangnan He.
\newblock Bias and debias in recommender system: A survey and future directions.
\newblock {\em ACM Trans. Inf. Syst.}, 41(3), feb 2023.

\bibitem{mansoury2020feedback}
Masoud Mansoury, Himan Abdollahpouri, Mykola Pechenizkiy, Bamshad Mobasher, and Robin Burke.
\newblock Feedback loop and bias amplification in recommender systems, 2020.

\bibitem{Imbalancedlearning}
Haibo He and Edwardo~A. Garcia.
\newblock Learning from imbalanced data.
\newblock {\em IEEE Transactions on Knowledge and Data Engineering}, 21(9):1263--1284, 2009.

\bibitem{fonseca2023tabular}
Joao Fonseca and Fernando Bacao.
\newblock Tabular and latent space synthetic data generation: a literature review.
\newblock {\em Journal of Big Data}, 10(1):115, 2023.

\bibitem{lu2024machine}
Yingzhou Lu, Minjie Shen, Huazheng Wang, Xiao Wang, Capucine van Rechem, Tianfan Fu, and Wenqi Wei.
\newblock Machine learning for synthetic data generation: A review, 2024.

\bibitem{smote_2002}
Kevin~W. Bowyer, Nitesh~V. Chawla, Lawrence~O. Hall, and W.~Philip Kegelmeyer.
\newblock {SMOTE:} synthetic minority over-sampling technique.
\newblock {\em CoRR}, abs/1106.1813, 2011.

\bibitem{CTGan_2019}
Lei Xu, Maria Skoularidou, Alfredo Cuesta-Infante, and Kalyan Veeramachaneni.
\newblock Modeling tabular data using conditional gan.
\newblock In H.~Wallach, H.~Larochelle, A.~Beygelzimer, F.~d\textquotesingle Alch\'{e}-Buc, E.~Fox, and R.~Garnett, editors, {\em Advances in Neural Information Processing Systems}, volume~32. Curran Associates, Inc., 2019.

\bibitem{SDV_2016}
Neha Patki, Roy Wedge, and Kalyan Veeramachaneni.
\newblock The synthetic data vault.
\newblock In {\em IEEE International Conference on Data Science and Advanced Analytics (DSAA)}, pages 399--410, Oct 2016.

\bibitem{TabDDPM_2022}
Akim Kotelnikov, Dmitry Baranchuk, Ivan Rubachev, and Artem Babenko.
\newblock Tabddpm: Modelling tabular data with diffusion models, 2022.

\bibitem{Khader2023}
Firas Khader, Gustav M\"{u}ller-Franzes, Soroosh Tayebi~Arasteh, Tianyu Han, Christoph Haarburger, Maximilian Schulze-Hagen, Philipp Schad, Sandy Engelhardt, Bettina Baeßler, Sebastian Foersch, Johannes Stegmaier, Christiane Kuhl, Sven Nebelung, Jakob~Nikolas Kather, and Daniel Truhn.
\newblock Denoising diffusion probabilistic models for 3d medical image generation.
\newblock {\em Scientific Reports}, 13(1), May 2023.

\bibitem{medical_fundus}
John~T. Guibas, Tejpal~S. Virdi, and Peter~S. Li.
\newblock Synthetic medical images from dual generative adversarial networks.
\newblock {\em CoRR}, abs/1709.01872, 2017.

\bibitem{medical_tumor_seg}
Hoo-Chang Shin, Neil~A. Tenenholtz, Jameson~K. Rogers, Christopher~G. Schwarz, Matthew~L. Senjem, Jeffrey~L. Gunter, Katherine~P. Andriole, and Mark Michalski.
\newblock Medical image synthesis for data augmentation and anonymization using generative adversarial networks.
\newblock In Ali Gooya, Orcun Goksel, Ipek Oguz, and Ninon Burgos, editors, {\em Simulation and Synthesis in Medical Imaging}, pages 1--11, Cham, 2018. Springer International Publishing.

\bibitem{Loecher2021}
Michael Loecher, Luigi~E. Perotti, and Daniel~B. Ennis.
\newblock Using synthetic data generation to train a cardiac motion tag tracking neural network.
\newblock {\em Medical Image Analysis}, 74:102223, December 2021.

\bibitem{Skandarani2023}
Youssef Skandarani, Pierre-Marc Jodoin, and Alain Lalande.
\newblock Gans for medical image synthesis: An empirical study.
\newblock {\em Journal of Imaging}, 9(3):69, March 2023.

\bibitem{Ho_2020}
Jonathan Ho, Ajay Jain, and Pieter Abbeel.
\newblock Denoising diffusion probabilistic models.
\newblock {\em CoRR}, abs/2006.11239, 2020.

\bibitem{diffusion_vs_gans}
Prafulla Dhariwal and Alexander Nichol.
\newblock Diffusion models beat gans on image synthesis.
\newblock In M.~Ranzato, A.~Beygelzimer, Y.~Dauphin, P.S. Liang, and J.~Wortman Vaughan, editors, {\em Advances in Neural Information Processing Systems}, volume~34, pages 8780--8794. Curran Associates, Inc., 2021.

\bibitem{UsmanAkbar2024}
Muhammad Usman~Akbar, Måns Larsson, Ida Blystad, and Anders Eklund.
\newblock Brain tumor segmentation using synthetic mr images - a comparison of gans and diffusion models.
\newblock {\em Scientific Data}, 11(1), February 2024.

\bibitem{villaiz2024diffusion}
Mario Villaiz{\'a}n-Vallelado, Matteo Salvatori, Carlos Segura, and Ioannis Arapakis.
\newblock Diffusion models for tabular data imputation and synthetic data generation, 2024.

\bibitem{Endres2022}
Markus Endres, Asha Mannarapotta~Venugopal, and Tung~Son Tran.
\newblock Synthetic data generation: A comparative study.
\newblock In {\em International Database Engineered Applications Symposium}, IDEAS’22. ACM, August 2022.

\bibitem{synrec_2018}
Manel Slokom.
\newblock Comparing recommender systems using synthetic data.
\newblock In {\em Proceedings of the 12th ACM Conference on Recommender Systems}, RecSys '18, page 548–552, New York, NY, USA, 2018. Association for Computing Machinery.

\bibitem{Noble2023RealisticBN}
Isaac Noble, Ivan Vendrov, Xinyuan Xu, and Deepak Ramachandran.
\newblock Realistic but non-identifiable synthetic user data generation.
\newblock In {\em EvalRS@KDD}, 2023.

\bibitem{fuxi_2021}
Jieming Zhu, Jinyang Liu, Shuai Yang, Qi~Zhang, and Xiuqiang He.
\newblock Open benchmarking for click-through rate prediction.
\newblock In Gianluca Demartini, Guido Zuccon, J.~Shane Culpepper, Zi~Huang, and Hanghang Tong, editors, {\em {CIKM} '21: The 30th {ACM} International Conference on Information and Knowledge Management, Virtual Event, Queensland, Australia, November 1 - 5, 2021}, pages 2759--2769. {ACM}, 2021.

\bibitem{fuxi_2022}
Jieming Zhu, Quanyu Dai, Liangcai Su, Rong Ma, Jinyang Liu, Guohao Cai, Xi~Xiao, and Rui Zhang.
\newblock {BARS:} towards open benchmarking for recommender systems.
\newblock In Enrique Amig{\'{o}}, Pablo Castells, Julio Gonzalo, Ben Carterette, J.~Shane Culpepper, and Gabriella Kazai, editors, {\em {SIGIR} '22: The 45th International {ACM} {SIGIR} Conference on Research and Development in Information Retrieval, Madrid, Spain, July 11 - 15, 2022}, pages 2912--2923. {ACM}, 2022.

\bibitem{deepfm}
Huifeng Guo, Ruiming Tang, Yunming Ye, Zhenguo Li, and Xiuqiang He.
\newblock Deepfm: A factorization-machine based neural network for ctr prediction, 2017.

\bibitem{masknet}
Zhiqiang Wang, Qingyun She, and Junlin Zhang.
\newblock Masknet: Introducing feature-wise multiplication to ctr ranking models by instance-guided mask, 2021.

\end{thebibliography}

\appendix

\end{document}